# Calibration plots for multistate risk predictions models: an overview and simulation comparing novel approaches

Alexander Pate[1], Matthew Sperrin[1,2], Richard D. Riley[3], Niels Peek[1,2], Tjeerd Van Staa[1], Jamie C. Sergeant[4,5], Mamas A. Mamas[6], Gregory Y. H. Lip[7,8], Martin O'Flaherty[9,11], Michael Barrowman[9], Iain Buchan[10,11], Glen P. Martin[1]

1. Centre for Health Informatics, Imaging and Data Science, Faculty of Biology, Medicine and Health, University of Manchester, Manchester Academic Health Science Centre, Manchester, United Kingdom

2. NIHR Manchester Biomedical Research Centre, University of Manchester, Manchester, UK

3. Institute of Applied Health Research, University of Birmingham, Birmingham, UK.

4. Centre for Epidemiology Versus Arthritis, Centre for Musculoskeletal Research, Manchester Academic Health Science Centre, University of Manchester, Manchester, UK

5. Centre for Biostatistics, Manchester Academic Health Science Centre, University of Manchester, Manchester, UK

6. Keele Cardiovascular Research Group, Keele University, Stoke-on-Trent, UK

7. Liverpool Centre for Cardiovascular Science at University of Liverpool, Liverpool John Moores University and Liverpool Heart & Chest Hospital, Liverpool, United Kingdom

8. Department of Clinical Medicine, Aalborg University, Aalborg, Denmark

9. NIHR Applied Research Collaboration NW Coast, University of Liverpool, Liverpool, UK

10. Institute of Population Health, Faculty of Health and Life Sciences, University of Liverpool, Liverpool, UK

11. Independent Researcher, Manchester, United Kingdom

# Abstract

## Introduction

There is currently no guidance on how to assess the calibration of multistate models used for risk prediction. We introduce several techniques that can be used to produce calibration plots for the transition probabilities of a multistate model, before assessing their performance in the presence of non-informative and informative censoring through a simulation.

## Methods

We studied pseudo-values based on the Aalen-Johansen estimator, binary logistic regression with inverse probability of censoring weights (BLR-IPCW), and multinomial logistic regression with inverse probability of censoring weights (MLR-IPCW). The MLR-IPCW approach results in a calibration scatter plot, providing extra insight about the calibration. We simulated data with varying levels of censoring and evaluated the ability of each method to estimate the calibration curve for a set of predicted transition probabilities. We also developed evaluated the calibration of a model predicting the incidence of cardiovascular disease, type 2 diabetes and chronic kidney disease among a cohort of patients derived from linked primary and secondary healthcare records.

## Results

The pseudo-value, BLR-IPCW and MLR-IPCW approaches give unbiased estimates of the calibration curves under non-informative censoring. These methods remained unbiased in the presence of informative censoring, unless the mechanism was strongly informative, with bias concentrated in the areas of predicted transition probabilities of low density.

## Conclusions

We recommend implementing either the pseudo-value or BLR-IPCW approaches to produce a calibration curve, combined with the MLR-IPCW approach to produce a calibration scatter plot, which provides additional information over either of the other methods.

# 1. Introduction

Clinical prediction models are used to guide clinical decision making in a variety of settings.[1–9] A multistate model may be used when the outcome of interest can potentially be reached via intermediate health states deemed important to include in the model, or may help improve prediction.[10] For example, when modelling the co-development of multiple long-term conditions (multimorbidity),[11–13] once an intermediate state has been developed (such as type 2 diabetes), this may increase the risk of developing further conditions (such as cardiovascular disease). Multistate models are used in a variety of clinical settings.[14–18]

Evaluation of the performance of a prediction model, through a validation study, is essential before it can be used in practice.[19] Model evaluation should routinely involve assessment of calibration and discrimination,[20–22] or other relevant techniques such as decision curve analysis,[23,24] measures of explained variation ($R^2$ statistics) or the Brier score.[25] The calibration of a model is evaluated by how well the estimated risks from the model agree with the observed event rates in a cohort of interest. Best practice for assessing calibration of a model predicting continuous, binary, polytomous or survival outcomes is well established,[20–22,26–36] with a (flexible) calibration plot being the preferred approach. Methods for evaluating the calibration of multi-state models are not well established. Titman and Sharples suggested comparing the risk of entry into an absorbing state (of a multistate model) with a Kaplan Meier estimate.[37] Spitoni et al.,[38] illustrated the estimation of prediction errors based on the Brier score, and Van Geloven et al.[29] reviewed methods for assessing calibration of a competing risks model (special case of a multistate model).

We reviewed a sample of studies published since 2017 involving the development of a multistate model for prediction. In studies that did assess model calibration, a wide range of approaches were taken. Postmus et al.,[39] compared the mean predicted risk of death (the primary state of interest) with mean observed survival calculated through a Kaplan Meier estimator, within subgroups defined by predicted risk. Beyer et al.,[40] analyse trial data, and compare the predicted value of overall survival from their multistate model with a Kaplan Meier estimate of overall survival within each trial arm. Deschepper et al.[41] compared predicted vs observed bed occupancy (state of interest), possible due to the lack of censoring. These approaches are valid but do not allow generation of calibration curves. Calibration was not assessed in most studies that were identified.[11–13,42–44] The lack of validation applied in practice may be due in part to the lack of guidance on model validation for multistate models.

In this paper, we detail how to assess calibration of the predicted transition probabilities from a multistate model using a variety of methods, focusing on methods that can produce calibration plots (Section 2). In Section 3, we compare the performance of such methods in a simulation study, and demonstrate their use in a clinical example in Section 4.

# 2. Methods for assessing calibration

## 2.1. Preliminaries

Let $X(t)$ be a multistate survival process with $K$ states, where $X(t) = k$ if an individual is in state $k$ at time $t$. Assume a multistate model has been developed to predict $X(t)$ using a set of predictor variables $\mathbf{Z}$ (e.g., one available in the literature, which we now aim to externally validate). Transition probabilities, $p_{j,k}(t_{start}, t_{end})$, are the probability of being in state $k$ at time $t_{end}$, when in state $j$ at time $t_{start}$. Transition probabilities are agnostic to the path taken between $j$ and $k$ and any number of intermediate states may have been visited (i.e., we are not only interested in 'direct' transitions, which is the case for competing risks). Let $t_{cens}$ be the censoring time, such that $X(t)$ is only observed for $t < t_{cens}$. Here, the aim is to assess the calibration of the estimated transition probabilities $\hat{p}_{j,k}(t_{start}, t_{end})$ in a cohort of interest.[45]

We consider cohorts where all individuals start in the same initial state. We restrict our assessment to the calibration of transition probabilities out of the initial state ($j = 1$), made at the start of follow up ($t_{start} = 0$). These two conditions are relevant to most clinical applications reported in the literature. Given this setting, we reduce the notation of the predicted transition probabilities to $\hat{p}_k(t)$. How to utilise landmarking[46,47] to evaluate calibration when these requirements do not hold is discussed in Section 4.

Four categories of calibration have previously been described:[35] 1) Mean calibration (or calibration-in-the-large) where "the average predicted risk is compared with the overall event rate". 2) Weak calibration where "on average, the model does not over- or underestimate risk, and does not give overly extreme (too close to 0 and 1) or modest (too close to disease prevalence or incidence) risk estimates". This is assessed using calibration intercept and slope. 3) Moderate calibration where the estimated risks correspond to observed proportions across the entire range of predicted risk. This would generally be assessed using flexible calibration curves,[27,28,32] where the observed event rate is plotted against the predicted risk. 4) Strong calibration requires "predicted risk corresponds to the observed proportion for every possible combination of predictor values", and is rarely considered in practice.

In sections 2.2 – 2.5, we detail a range of methods that can be used to assess the calibration of the transition probabilities out of the initial state in a multistate model. We focus on calibration plots, but details are also provided on how to assess mean and weak calibration.

## 2.2. Aalen-Johansen estimator

The Aalen-Johansen estimator[48] is an estimator for the transition probability matrix of a non-homogeneous Markov process (a generalisation of the Kaplan Meier estimator to multistate survival data). This can be used to estimate the observed risk of each state in the cohort of interest, $obs_k^{AJ}(t)$. The mstate[49] package can be used to derive the Aalen-Johansen estimator.

Mean calibration using this estimator is assessed by comparing the mean predicted transition probability with the observed risk for each outcome category:

$$\left(\sum_{i=1}^{n} obs_k^{AJ}(t) - \hat{p}_k^i(t)\right)/n,$$

where $\hat{p}_k^i(t)$ is the predicted transition probability for individual $i$. This estimator makes the assumptive of non-informative censoring. In the presence of informative censoring, we recommend calculating the Aalen-Johansen estimator within sub-groups of individuals defined by the predicted transition probability of each state and taking the average. Specifically, for each category $k$, order individuals by predicted transition probability of category $k$, split into equally sized groups, within each group calculate the Aalen-Johansen estimator, take the average and subtract the mean predicted transition probability in the cohort. This relaxes the non-informative censoring assumption, to non-informative censoring after conditioning on the predicted transition probabilities. Moderate calibration is assessed by comparing the Aalen-Johansen estimator and average predicted transition probability within each sub-group. Importantly, the Aalen-Johansen approach does not allow the estimation of calibration curves and therefore the approach is not recommended. However, knowledge of this estimator is required to motivate the pseudo-value approach given in Section 2.3.

Finally, a discussion of the Markov assumption is required. In general, the Aalen-Johansen estimator is only a consistent estimator under the Markov assumption for the multistate process $X(t)$, which means at all times $t$ *"given the present state and the event history of a patient, the next state to be visited and the time at which this will occur will only depend on the present state"*.[10] However, it can be shown through the work of Datta and Satten[50] and Glidden,[51] that in this specific setting (all individuals start in the same initial state), that the Aalen-Johansen estimator gives a consistent estimate of the transition probabilities out of this initial state. Therefore Estimators which do not depend on the Markov assumption[52–55] are not required here. A full line of reasoning is given in Supplementary Material part 1.

## 2.3. Pseudo-values

Pseudo-values enable estimation of estimands that can be expressed as an expectation when analysing censored survival data.[56] The pseudo-value itself denotes the contribution of each individual to the expectation that is being estimated. If $\hat{\theta}$ is such an estimator, the pseudo-value for individual $i$ is defined as:

$$\hat{\theta}^i = n * \hat{\theta} - (n-1) * \hat{\theta}^{-i},$$

where $\hat{\theta}^{-i}$ is the estimator $\hat{\theta}$ calculated in the cohort with individual $i$ removed. For more details on pseudo-observations, and the required properties of the estimator $\hat{\theta}$ and estimand $\theta$, see the work of Andersen and Perme.[56]

Pseudo-values for the transition probabilities from a multistate model can be calculated using the Aalen-Johansen estimator.[57] Let $\hat{\theta}^i(t)$ be the vector of pseudo-values for individual $i$ at time $t$, where $\hat{\theta}^i_k(t)$ is the pseudo-value for the transition to state $k$. Moderate calibration for each state can be evaluated by regressing the pseudo-values on the predicted transition probabilities using a loess smoother:

$$\hat{\theta}^i_k(t) = loess\left(\hat{p}^i_k(t)\right). \tag{1}$$

Other models and link functions could also be used.[56] Observed probabilities $\widehat{obs}_k^{PV,i}(t)$ can then be obtained by generating fitted values for each individual in the validation cohort from model (1). The set of points $\{\hat{p}^i_k(t), \widehat{obs}_k^{PV,i}(t)\}$ gives the calibration plot for state $k$.

The estimates of calibration derived using pseudo-values are based on the same assumptions as the underlying estimator $\hat{\theta}$. For example, when $\hat{\theta}$ is the Aalen-Johansen estimator, censoring must be non-informative. However, if the censoring mechanism and outcome are conditionally independent given some variables $Z$, then the pseudo-values will be valid if calculated within subgroups defined by $Z$.[56] We therefore suggest calculating pseudo-values within subgroups defined by the predicted transition probabilities. Specifically, for each state $k$, order individuals by $\hat{p}^i_k(t)$ and split into groups (say deciles or ventiles). Within each group, calculate $\hat{\theta}^i$ using the Aalen-Johansen estimator and extract $\hat{\theta}^i_k(t)$. Note that the ordering of individuals may be different for each state $k$. All individuals are then recombined into a single dataset when fitting model (1).

The mean calibration for each outcome category can be calculated as the average difference between the pseudo-values and predicted transition probabilities:

$$\left(\sum_{i=1}^{n} \hat{\theta}^i_k(t) - \hat{p}^i_k(t)\right)/n$$

To assess weak calibration, fit the following linear regression model:

$$\hat{\theta}_k^i(t) = \beta * \hat{p}_k^i(t),$$

and report the calibration slope $\hat{\beta}$.

### 2.4. Binary logistic recalibration using inverse probability of censoring weights (BLR-IPCW)

Barrowman et al.,[15] first proposed evaluating the calibration of $\hat{p}_k(t)$ by estimating calibration intercepts and slopes (at a given time point $t$) using binary logistic regression calibration techniques[58,59] and inverse probability of censoring weights. We extend the approach to assess moderate calibration using calibration curves.

Let $I_k(t)$ be an indicator variable indicating whether an individual is in state $k$ at time $t$ (for all individuals uncensored at time $t$). The indicator $I_k(t)$ can be modelled in the validation cohort using a loess smoother[32]:

$$I_k(t) = loess\left(\hat{p}_k^i(t)\right). \quad (2)$$

Fitted values $\widehat{obs}_k^{BLR,i}(t)$ can then be generated for each individual in the validation cohort using Model (2), and the set of points $\{\hat{p}_k^i(t), \widehat{obs}_k^{BLR,i}(t)\}$ results in the calibration plot for state $k$. Other models could be used to model the relationship between $I_k(t)$ and $\hat{p}_k(t)$, such as a binary logistic regression model (after applying logit transformation to $\hat{p}_k^i(t)$) with multiple fractional polynomials[60] or restricted cubic splines.[20]

In the absence of censoring, the above approach would allow us to estimate the estimand of interest, which is the observed event probabilities for state $k$ among individuals with a predicted transition probability of $\hat{p}_k(t)$:

$$P(X(t) = k|\hat{p}_k(t))$$

However, in the presence of censoring this can only be done in the subgroup of the population which is uncensored at time $t$. The assessment of calibration will be biased by censoring (induces selection bias) unless the censoring mechanism is independent from everything else. The calibration Model (2) must therefore be fitted using data reweighted according to inverse probability of censoring weights[25,61] (IPCW) to obtain valid estimates. The weights are calculated as the inverse of the probability that an individual has not been censored by time point $t$:

$$w_t = \frac{1}{P(t_{cens} > t|\boldsymbol{Z}, \boldsymbol{X(t)})},$$

where $\boldsymbol{Z}$ is a vector of baseline covariates, and $\boldsymbol{X(t)}$ denotes the position of the multistate process at time $t$, and information on any previous transition times. The inclusion of $\boldsymbol{X(t)}$ reflects that if an individual enters an absorbing state at time $t_{abs} < t$, the weight for this individual is the probability of being uncensored at time $t_{abs}$, as opposed to time $t$. Specifically: $P(t_{cens} > t|Z, \boldsymbol{X(t)}) = P(t_{cens} > t_{abs}|Z)$ if $X(t)$ is in an absorbing state, and $P(t_{cens} > t|Z, \boldsymbol{X(t)}) = P(t_{cens} > t|Z)$ otherwise. In the presence of an absorbing state, it is therefore imperative to apply IPCWs even when censoring is non-informative. These weights should be estimated from the data by modelling the time until censoring occurs.

To assess mean calibration, one would fit the following model:

$$logit[I_k(t)] = \alpha + \beta * logit(\hat{p}_k(t)), \quad (3)$$

fixing $\beta$ to 1. The estimate $\hat{\alpha}$ (calibration intercept) gives an assessment of mean calibration on the logit scale. We recommend the following steps to transform this onto the probability scale, in line with those from Sections 2.2 and 2.3. Calculate observed event probabilities by generating fitted values from Model (3):

$$\widehat{obs}_k^{BLR-mean,i}(t) = logit^{-1}\left(\hat{\alpha} + logit\left(\hat{p}_k^i(t)\right)\right)$$

The mean calibration for state $k$ can then be estimated as the mean difference between the observed event probabilities and the predicted transition probabilities:

$$\left(\sum_{i=1}^{n} \widehat{obs}_k^{BLR-mean,i}(t) - \hat{p}_k^i(t)\right)/n \, .$$

To assess weak calibration, one fits Model (2) with no constraints to calculate the calibration slope, $\hat{\beta}$, where values less than 1 indicate overfitting.

## 2.5. Nominal recalibration framework using multinomial logistic regression and inverse probability of censoring weights (MLR-IPCW)

Barrowman et al., [15] also proposed calculating calibration intercepts and slopes using the nominal calibration framework of Van Hoorde et al.,[33] developed for assessing the calibration of polytomous outcomes to model the observed event probabilities of all outcome categories simultaneously. We extend this to assess moderate calibration by adapting the spline-based extension of the nominal calibration framework.[34]

Let $I_X(t)$ be a polytomous variable taking values in $\{1, \ldots, K\}$, such that $I_X(t) = k$ if an individual is in state $k$ at time $t$. For each individual $i$ in the validation cohort, calculate the log-ratio of the probability of each state ($k \neq 1$) relative to state 1:

$$\widehat{LP}_k^i = \ln\left[\frac{\hat{p}_k^i(t)}{\hat{p}_1^i(t)}\right]$$

The outcome $I_X(t)$ is then modelled using a multinomial logistic regression with a vector spline smoother:

$$\ln\left[\frac{P(I_X(t) = k)}{P(I_X(t) = 1)}\right] = \alpha_k + \sum_{h=2}^{K} \beta_{k,h} * s_k(\widehat{LP}_h), \quad (4)$$

for $k > 1$. Observed event probabilities values $\widehat{obs}_k^{MLR,i}(t)$ can then be obtained by generating fitted values for each individual in the validation cohort from Model (4) using the following formulae:

$$\widehat{obs}_1^{MLR,i}(t) = \frac{1}{1 + \sum_{l=2}^{K}(\hat{\alpha}_l + \sum_{h=2}^{K} \hat{\beta}_{l,h} * s_l(\widehat{LP}_h))}$$

$$\widehat{obs}_k^{MLR,i}(t) = \frac{exp(\hat{\alpha}_k + \sum_{h=2}^{K} \hat{\beta}_{k,h} * s_k(\widehat{LP}_h))}{1 + \sum_{l=2}^{K}(\hat{\alpha}_l + \sum_{h=2}^{K} \hat{\beta}_{l,h} * s_l(\widehat{LP}_h))}$$

where $\widehat{obs}_k^{MLR,i}(t)$ is defined for $k > 1$. The set of points $\{\hat{p}_k^i(t), \widehat{obs}_k^{MLR,i}(t)\}$ then gives the calibration plot for state $k$. Alike to the methods in Section 2.4, this process can only be done in the sub-cohort of individuals uncensored at time $t$, and therefore IPCW must be applied to obtain valid estimates of calibration. Weights are calculated the same way as in section 2.4. The VGAM package[62] can be used to fit these multinomial logistic regression models, and generate observed event probabilities (fitted values). For more details on the vector spline smoother see Van Hoorde et al.[34]

Note that unlike MLR-IPCW, the BLR-IPCW and pseudo-value approaches ignore the multistate aspect of the data and treat each outcome category with a 'one vs all' approach. This means the observed event probabilities for an individual will only sum to 1 when using the MLR-IPCW approach. Furthermore, the calibration plots for MLR-IPCW are a scatter plot, reflecting the fact that individuals with the same predicted transition probability of state $k$, may have different predicted transition probabilities of the other states, and in turn different observed event probabilities. This approach can therefore provide more insight into the calibration of a model over the BLR-IPCW and pseudo-value approaches.

To assess mean calibration, one would fit the following multinomial logistic regression model in the validation cohort:

$$\ln\left[\frac{P(I_X(t) = k)}{P(I_X(t) = 1)}\right] = \alpha_k + \beta_k * \widehat{LP}_k \, , \quad (5)$$

where the coefficients $\beta_k$ are fixed to 1. The estimated $\hat{\alpha}_k$ constitute the calibration intercepts from the calibration framework of Van Hoorde et al.[33] To get this estimate of mean calibration onto the probability scale, we again calculate observed event probabilities by generating fitted values from Model (4) for each individual $i$ in the validation cohort using the following formulae:

$$\widehat{obs}_1^{MLR-mean,i}(t) = \frac{1}{1 + \sum_{l=2}^{K} \exp(\hat{\alpha}_l + \widehat{LP}_l^i)}$$

$$\widehat{obs}_k^{MLR-mean,i}(t) = \frac{\exp(\hat{\alpha}_k + \widehat{LP}_k^i)}{1 + \sum_{l=2}^{K} \exp(\hat{\alpha}_l + \widehat{LP}_l^i)}$$

The mean calibration for state $k$ can then be calculated as the mean difference between the observed event probabilities and the predicted transition probabilities:

$$\left(\sum_{i=1}^{n} \widehat{obs}_k^{MLR-mean,i}(t) - \hat{p}_k^i(t)\right)/n \, .$$

To assess weak calibration, fit Model (5) with no constraints and calculate the calibration slopes $\hat{\beta}_k$.

# 3. Simulation

The methods are presented using the Aims, Data generating mechanisms, Estimands, Methods and Performance (ADEMP) structure.[63] All code for running the simulation is available on GitHub.[64]

## 3.1. Aims

The primary aim was to compare the bias (and variability) of each of the calibration methods outlined in Section 2 when the assumption of non-informative censoring does and does not hold.

## 3.2. Data generating mechanisms

### 3.2.1. Underlying structures

Data was generated using the multistate model structure in Figure 1. This was designed to mimic the clinical setting from Putter et al.,[16] predicting local recurrence (state 2), distant metastasis (state 3), local recurrence + distant metastasis (state 4), and death (state 5), after surgery (state 1) in patients with early stage breast cancer.

We first generated two baseline predictor variables $Z_1$ and $Z_2$, both drawn from a standard Gaussian distribution. The survival times for transition $j \rightarrow k$ were simulated from an exponential distribution with hazard $exp(0.5 * Z_1 - 0.5 * Z_2) * \lambda_{j,k}$. The predictor $Z_1$ represents a variable associated with higher risk of remission and/or death, and $Z_2$ represents a variable associated with lower risk of remission and/or death. The baseline hazard for each transition ($\lambda_{j,k}$) was chosen so that the 7-year survival matched that of the cause-specific hazard reported by Putter et al.[16] The targeted values for each transition (and corresponding values of $\lambda_{j,k}$) are provided in Table 1. 7-year survival was targeted as this was the maximum follow up observed for all transitions. The censoring time was generated from an exponential survival distribution with hazard $exp(\beta_{cens,1} * Z_1 + \beta_{cens,2} * Z_2) * \lambda_C$. The value of $\lambda_C = 5005$ was chosen to induce a censoring probability of 0.4 over 7 years for individuals with $Z_1 = Z_2 = 0$. An uninformative censoring scenario (UIC) was induced by setting $\beta_{cens,1} = \beta_{cens,2} = 0$. A weakly informative censoring scenario (WIC) was induced by setting $\beta_{cens,1} = 0.25$, $\beta_{cens,2} = -0.25$. A strongly informative censoring scenario (SIC) was induced by setting $\beta_{cens,1} = 1$, $\beta_{cens,2} = -1$.

*[INSERT Figure 1: Data generating mechanism for the simulation]*

*Table 1: Targeted 7-year survival and corresponding baseline hazards $\lambda_{j,k}$ used to generate the data in data generating mechanism 1*

| Transition | 1 → 2 | 1 → 3 | 1 → 5 | 2 → 4 | 2 → 5 | 3 → 4 | 3 → 5 | 4 → 5 |
|---|---|---|---|---|---|---|---|---|
| Targeted survival probability | 0.9 | 0.8 | 0.99 | 0.55 | 0.95 | 0.7 | 0.15 | 0.05 |
| $\lambda_{j,k}$ | 24267 | 11458 | 254394 | 4277 | 49856 | 7168 | 1348 | 853 |

### 3.2.2. Process for simulation

Large validation sample ($n = 200{,}000$) and small validation sample ($n = 3000$, $n = 1500$) simulations were carried out for each scenario (NIC, WIC and SIC). Bias (bias and variance) of the calibration assessment was evaluated in the large (small) validation sample analysis. A sample size of 3000 was chosen to match the sample size used in the work of Putter et al.[16] A sample size of 1500 was chosen to test small sample performance even further. Sample size criteria do not yet exist for multistate models,[65] but would be dependent on the event rates of all the possible transitions. With this data generating mechanism some of the transitions are rare (approximately 2% of individuals are in state 4 after 7 years). We therefore believe 1,500 should be considered a very low sample size.

The large validation sample simulation followed the following steps for each scenario:

1. Generate a validation dataset of size $n = 200{,}000$ according to the relevant data generating mechanism.

2. Calculate the true transition probabilities for each individual at 7 years follow-up, $p_k(7)$.

3. Define $\hat{p}_{k,perf} = p_k$ as a perfectly calibrated estimate of the transition probabilities. Define $\hat{p}_{k,over} = logit^{-1}(logit(p_k) + 0.5)$ and $\hat{p}_{k,under} = logit^{-1}(logit(p_k) - 0.5)$ as miss-calibrated over and under estimates of the transition probabilities.

4. Assess mean and moderate calibration of $\hat{p}_{k,perf}$, $\hat{p}_{k,over}$ and $\hat{p}_{k,under}$ in the dataset using each of the calibration approaches (section 3.4).

5. Compare calibration to true calibration (estimand, Section 3.3) for moderate and mean calibration.

For the small sample simulation, a cohort of size 1,000,000 was generated. Cohorts of size 3000 or 1500 were then sampled at random (without replacement) and Steps 2 – 5 were implemented. For Step 5 mean calibration was assessed. This process was repeated 1,000 times. This represents the process of validating the model in a cohort sampled randomly from the population of interest.

### 3.3. Estimands and other targets

The estimand of interest is the calibration of the predicted transition probabilities defined in the previous section ($\hat{p}_{k,perf}$, $\hat{p}_{k,over}$ and $\hat{p}_{k,under}$). Let $\hat{p}_k^i$ represent any of $\hat{p}_{k,perf}$, $\hat{p}_{k,over}$ and $\hat{p}_{k,under}$. Let $p_k^i$ be the 'true' transition probabilities from state 1 to state $k$ for individual $i$ used in the data generating mechanism. These can be calculated from the specified cause specific hazards in section 3.2.1 following Putter et al.[10] The exact process and formulae for doing so is provided in Supplementary Material 1. When assessing moderate calibration, the estimand is the set of points $\{\hat{p}_k^i, p_k^i\}$. When assessing mean calibration, the estimand of interest is the average difference between the predicted transition probabilities and true transition probabilities:

$$\left(\sum_{i=1}^{n} p_k^i - \hat{p}_k^i\right)/n.$$

Note that the estimand is a function of both the 'true' transition probabilities used to generate the data and the predicted transition probabilities. In practice the predicted transition probabilities ($\hat{p}_k^i$) would be estimated by fitting a multistate model, however in this simulation they have been defined deterministically. The manner in which $\hat{p}_{k,over}$ and $\hat{p}_{k,under}$ were generated will result in a non-linear calibration curves, desirable when assessing the ability of the proposed methods to evaluate moderate calibration.

### 3.4. Methods for analysis

When assessing moderate calibration, we compared the BLR-IPCW, MLR-IPCW and pseudo-values based on Aalen-Johansen estimator. The Aalen-Johansen estimator was not considered given it cannot estimate the estimand ($\{\hat{p}_k^i, p_k^i\}$). When assessing mean calibration we compared BLR-IPCW, MLR-IPCW and the Aalen-Johansen estimator. The pseudo-value approach was not considered as the only advantage of the pseudo-value approach over the Aalen-Johansen estimator is that it enables the estimation of calibration curves. For the pseudo-value and Aalen-Johansen approaches, individuals were grouped into 20 groups based on their risk of state $k$ in order to assess calibration, as outlined in Sections 2.2 and 2.3. We apply this extra step in the NIC scenario, even though this would not be required in practice if the censoring mechanism is known to be non-informative.

We also conducted sensitivity analyses to assess the sensitivity of BLR-IPCW and MLR-IPCW to the calculation of the weights. For the main implementation of BLR-IPCW and MLR-IPCW, the weights were estimated from the data using a perfectly specified Cox model which adjusted for $Z_1$ and $Z_2$. A second analysis was carried out using a misspecified Cox model for the time until censored which didn't adjust for $Z_1$ and $Z_2$. A third analysis used the true weights, where the probability of being censored was calculated based off the data generating mechanism. A fourth analysis did not apply weights at all.

## 3.5. Performance metrics

In the large sample simulation, for moderate calibration bias was assessed by graphically comparing the estimated calibration curves from each method vs the true calibration curve represented by the set of points $\{\hat{p}_k^i, p_k^i\}$. For mean calibration the bias and standard error of the estimated mean calibration was reported for each method. Standard errors were calculated using bootstrapping, although we do not report the standard error for the Aalen-Johansen approach for computational reasons, given it is expected to be very small anyway. For the small sample simulation, we report the median bias and 2.5 – 97.5 percentile range (across the 1000 simulation iterations) in the bias of the mean calibration estimate.

## 3.6. Results

### 3.6.1. Large sample simulation: Moderate calibration

Figure 2 contains the calibration plots of the perfectly calibrated ($\hat{p}_{k,perf}$), over predicting ($\hat{p}_{k,over}$) and under predicting transition probabilities ($\hat{p}_{k,under}$) in scenario NIC (blue lines), against the calibration plot produced by using BLR-IPCW to assess calibration. Equivalent plots for the pseudo-value and MLR-IPCW approaches are provided in the supplementary material (Figures S1 and S2). These Figures showcase the ability of each method to appropriately assess non-linear changes in calibration over the distribution of predicted transition probabilities. All methods provide an unbiased assessment of the calibration when censoring is non-informative. There are some minor deviations from the true calibration for states 2, 3 and 4, for which the predicted transition probabilities are much smaller, highlighting a bigger sample size is required to assess moderate calibration of rarer states. Variability in calibration estimates is considered in the small validation sample simulation.

Figures 3 and 4 contain the moderate calibration plots of the perfectly calibrated transition probabilities for scenarios WIC and SIC. As the strength of informative censoring increases, this introduces bias into all the calibration methods. However, even when informative censoring is at its highest (Figure 4), BLR-IPCW, MLR-IPCW and pseudo-value all provide a predominately unbiased assessment of calibration. The rug plots indicate that the bias only occurs over the range of predicted transition probabilities where data points are sparse. Comparing the BLR-IPCW and pseudo-value approaches (the calibration curves) in the SIC scenario (Figure 4), shows each method gives a better evaluation of calibration for different states. Equivalent plots for the over and under predicting transition probabilities (Figures S3 – S6) lead to the same conclusions.

The variance in the points in the MLR-IPCW scatter plot for each state is very small. This is due to a lack of variability in the predicted transition probabilities of the other states. The added information provided by this approach is therefore not evident from the simulation. We refer to the clinical example (Section 4) for an illustration of the MLR-IPCW approach where the scatter plot is more varied, and discuss the benefits of this approach in more detail there. It also appears there is a small number of data points where calibration is biased over dense regions of predicted transition probability (e.g. state 4 in Figure 4), in contrary to the arguments in the previous paragraph. However, this is due to the multidimensional nature of these calibration plots. These points represent a small group of individuals who have the same predicted transition probability of state 4 as many other individuals, but have predicted transition probabilities of the other states which few people do. Therefore this is in fact a sparse area of predicted transition probability with respect to the other states.

We tested the sensitivity of the BLR-IPCW and MLR-IPCW approaches to miss-specification of the censoring distribution (Figures S7 – S24). This resulted in a small increase in the bias of these methods, however the calibration was still unbiased over a large range of the predicted transition probabilities, indicating some protection against misspecification of the censoring distribution.

***[INSERT Figure 2: Moderate calibration of BLR for perfectly calibrated transition probabilities ($\widehat{p}_{k,perf}$), over predicting transition probabilities ($\widehat{p}_{k,over}$) and under predicting transition probabilities ($\widehat{p}_{k,under}$) in scenario NIC, large sample analysis.]***

***[INSERT Figure 3: Moderate calibration of each method for perfectly calibrated transition probabilities ($\widehat{p}_{k,perf}$) in scenario WIC, large sample analysis.]***

***[INSERT Figure 4: Moderate calibration of each method for perfectly calibrated transition probabilities ($\widehat{p}_{k,perf}$) in scenario SIC, large sample analysis.]***

### 3.6.2. Large sample simulation: mean calibration

When there is no informative censoring (NIC), all methods give an unbiased estimate of the mean calibration (Figure 5). WIC starts to introduce bias into the mean calibration of the BLR-IPCW and MLR-IPCW approaches for states 1 and 5. When large levels of informative censoring are introduced (SIC) the bias increases further, whereas the Aalen-Johansen approach remained unbiased. Comparing this to the moderate calibration plots (Figures 4, S5 and S6), indicates the bias in the mean calibration is driven by a small number of outliers. Bias is only found for states 1 and 5 because increasing $Z_1$ or $Z_2$ only has a minor impact on the probability of being in states 2, 3 and 4. A change in $Z_1$ or $Z_2$ will increase or decrease the hazard for transitions both into and out of intermediate states. For example, an increase in $Z_1$ increases the rate of entry into state 2, but also increases the rate at which individuals leave state 2, having a net zero effect. This highlights that the issue of informative censoring may be present for some states but not others. We tested the sensitivity of the BLR-IPCW and MLR-IPCW approaches to misspecification of the weights (Figures S25 and S26). Misspecification of the weights had a minor effect on the bias, however implementing either approach without weights greatly increased the bias.

### 3.6.3. Small sample simulation: mean calibration

The results from the small sample simulation (N = 3000) are presented in Figure 6. In scenarios NIC and WIC all methods performed very similarly. For SIC there were some differences in performance for prediction of states 1 and 5, the states where informative censoring was at its strongest. The magnitude of bias was similar for AJ, BLR-IPCW and MLR-IPCW (although sometimes in opposite directions), but the variation in estimates was much bigger for the AJ approach. Reducing the number of groups patients are categorised into before evaluating calibration only had a minor impact on reducing this variation (Figures S27 and S28). Corresponding figures for N = 1500 (Figures 29 and S30) lead to similar conclusions.

***[INSERT Figure 5: Bias (CI) of each method for assessing mean calibration, large sample analysis.]***

***[INSERT Figure 6: Median bias and 2.5 – 97.5 percentile range in bias of each method for assessing mean calibration, small sample analysis.]***

# 4. Clinical example

## 4.1. Aim and setting

We aimed to demonstrate the application of each calibration method to typical clinical data and contrast the levels of information provided by each method. We developed an illustrative (not intended for clinical use) multistate model to predict the 10-year risk of co-existence of three long-term conditions, cardiovascular disease, type 2 diabetes and chronic kidney disease, in healthy individuals – a common and important multimorbidity state in which the develop of one condition increases the probability of the development of the other conditions. The clinical and economic impact of multimorbidity is high and rising in many parts of the world,[66–75] and preventing it is a priority for health systems.

## 4.2. Methods

**Data source:** Data from the Clinical Practice Research Datalink (CPRD) Aurum, a primary care dataset containing data from general practices with the EMIS Web® computer systems in England and Northern Ireland, was used is this study. CPRD Aurum contains > 39 million historical patients, and > 13 million currently registered. It is representative of the English population in terms of age, gender, geographical spread and deprivation (as of 2019).[76] These records were linked to Hospital Episode Statistics (HES) and deaths data from the Office for National Statistics (ONS).

Start of follow up was defined as the maximum of date turned age 65, 1st Jan 2000, and date of 1 year of up to standard registration in the database. Inclusion in the cohort was dependent on these three conditions being met. Patients were excluded if they had any of the outcomes prior to start of follow up. Censoring was defined as last date of data collection for the practice, or date transferred out of practice, which should be an uninformative censoring mechanism.

We extracted the transition times for the outcomes of incident cardiovascular disease, type 2 diabetes and chronic kidney disease (stage 3, 4 or 5), multimorbidity and death (Figure 7). There was a transition from every possible state to death. Outcome events were identified through the primary care, secondary care and death record data sources. Baseline data was extracted solely from the primary care record. We extracted: age, gender, systolic blood pressure, body mass index, total cholesterol/high density lipoprotein ratio, smoking status, index of multiple deprivation, and history of hypertension, depression and alcohol misuse. Code lists for all variables are available in Supplementary Material 1. To keep the scope of this example manageable, we wanted to focus on complete case datasets, however data were missing for systolic blood pressure, body mass index, cholesterol/high density lipoprotein and smoking status. We therefore generated a pseudo-complete case dataset by imputing missing values using a single stochastic imputation (achieved through a single multiple imputation chain of length 20, using all other variables as predictors).

*[INSERT Figure 7: A multistate model for prediction of cardiovascular disease, type 2 diabetes and chronic kidney disease]*

**Model development and validation:** Two development datasets (N = 5,000 and N = 100,000) and one validation dataset (N = 100,000) were sampled at random from the overall cohort. Multistate models were developed in both development datasets following the approaches of Putter et al.,[10] and de Wreede et al.,[49] taking a clock forward approach. Cox proportional hazard models adjusting for all baseline variables were fitted for each cause-specific hazard depicted in Figure 7. Note it was also possible that individuals were diagnosed with two conditions on the same date (i.e. transition from state 1 straight to state 5). However, we do not believe these transitions to be representative of the true underling disease process (long term conditions were likely developed at different times and only recorded in database once the patient engaged with the health system). The number of individuals who made these transitions was also small. We therefore modelled the cause-specific hazard for these transitions using a Kaplan Meier estimator and make no adjustment for baseline predictors.

After model development, transition probabilities were estimated for every patient in the validation cohort following the process of Putter et al.,[10] and de Wreede et al.[49] The censoring distribution was estimated using a Cox proportional hazards model adjusting for all baseline covariates. This model was used to estimate each individuals probability of being censored at 10-years, which were converted into IPCWs (capped at a maximum weight of $10^{77}$). For individuals who entered the absorbing death state, the probability of being uncensored at the time of death was used. Mean and moderate calibration was then assessed using the Aalen-Johansen, BLR-IPCW, MLR-IPCW and pseudo-values (based on the Aalen-Johansen estimator) approaches.

### 4.3. Results

The number of transitions over the entire course of follow up in the development (N = 5,000) and validation (N = 100,000) datasets are given in supplementary Table S1. The moderate calibration according to each method for N = 5,000 development sample are presented in Figure 8. Focusing on the pseudo-value and BLR-IPCW calibration curves, calibration of the transition probabilities of being in state 2 (CVD) is the worst, under predicting for lower risks and over predicting higher risks (possibly a sign of overfitting). A similar pattern is seen for transition probabilities into state 4 (CKD), although it is less extreme. There is over prediction of the transition probabilities into state 3 (CKD) and state 5 (multimorbidity) only at the highest predictions. Calibration of the transition probabilities into states 1 (healthy) and 6 (death) are the strongest.

The MLR-IPCW calibration scatter plot reflects quite a lot of variation in the calibration of the transition probabilities into states 2 and 4. The interpretation of this is that even for individuals with predicted transition probabilities which are deemed to be 'well calibrated' according to the BLR-IPCW or pseudo-value plots, there may be considerable miscalibration depending on their predicted transition probabilities for the other states. This contrasts with the calibration scatter plots for the model developed on the N = 100,000 sample (Figure S31). While the plots for BLR-IPCW and pseudo-values look similar in Figures 8 and S31, the difference in the calibration scatter plots produced by MLR-IPCW is notable. This evidence would lead to the conclusions that the N = 100,000 is a better calibrated model, a conclusion that may not have been reached looking only at the BLR-IPCW and pseudo-value plots.

*[INSERT Figure 8: Moderate calibration according to each method (development sample size = 5000).]*

## 5. Discussion

We have detailed a range of techniques for evaluating the calibration of the transition probabilities out of the initial state of a multistate model. The BLR-IPCW, MLR-IPCW and pseudo-value approaches each yielded unbiased calibration curves (or scatter plots for MLR-IPCW) under non-informative censoring. The calibration plots showed relatively small bias even in the presence of strong levels of informative censoring. At small sample sizes all methods had similar levels of bias when assessing mean calibration, however variability in the Aalen-Johansen estimator was larger than for BLR-IPCW and MLR-IPCW. We recommend assessing calibration using one of either BLR-IPCW or pseudo-value approach, both of which produce a smoothed calibration curve, alongside MLR-IPCW, which provides a deeper evaluation of calibration through a scatter plot.

The BLR-IPCW and pseudo-value approaches produce curves because they assess the calibration of the transition probabilities into each state vs not being in that state, taking a "one vs all" approach. On the other hand, MLR-IPCW considers the probability of being in each state simultaneously, providing information on whether the observed event rate for state $k$ (state of interest) differs for individuals with the same risk of being in state $k$, but different risks of being in states $\neq k$. To illustrate this, suppose we have a three-state illness-death model, where individuals with a 50% risk of being in the 'illness' state at time $t$ either have a 50% risk of being in the 'healthy' state (group A), or have a 50% risk of being in the 'death' state (group B), and that these groups are the same size. Now suppose that the model overpredicts the risk of being in the 'illness' state at time $t$ for group A, and underpredicts for group B by the same amount. Calibration of the transition probabilities into the 'illness' state is therefore poor for all these individuals. However, the BLR-IPCW and pseudo-value approach would find the model to be well calibrated for these individuals as the event rate would equal the average predicted risk across all individuals in both groups A and B. This type of miscalibration would be identified by the MLR-IPCW approach (see Van Hoorde et al.,[34] for further discussion of this approach). Note that that confidence intervals can be calculated for the calibration curves using bootstrapping, however, assessing the variability in the scatter plot produced by MLR-IPCW is less tractable.

A benefit of the BLR-IPCW and MLR-IPCW approaches over the pseudo-value approach based on the Aalen-Johansen estimator is that they are cross-sectional and make no assumptions about the time scale or whether the Markov assumption holds. For example, in this study we have assumed the time scale in all models to be clock-forward,[10] however in practice the analyst may want to assess the calibration of a model developed under a clock-reset framework (where time resets to 0 after entry into a new state). This is also known as a semi-Markov or Markov renewal model, and the Markov assumption is violated. While we argued that the Aalen-Johansen approach was valid even when the Markov assumption was violated under our assumptions (section 2.2), this would not be the case if some individuals started in subsequent states, or if we were interested in assessing calibration of transition probabilities out of subsequent states. Biologically, end-organ susceptibility to common risk factors such as obesity varies in populations, where one individual may suffer kidney damage from high blood pressure, another may develop diabetes before kidney disease. In this case, other estimators do exist that could replace Aalen-Johansen, such as the Landmark Aalen-Johansen estimator,[52] or estimators designed specifically for Markov renewal processes.[78,79] While these estimators can replace the Aalen-Johansen estimator when calculating the pseudo-values, the added benefit of BLR-IPCW and MLR-IPCW is that they can be applied ubiquitously to validate a multistate model developed with any range of assumptions or structure. A second advantage is that the pseudo-value approach takes considerably longer to implement than the BLR-IPCW or MLR-IPCW approaches from a computational perspective, which impacts the ability to calculate confidence intervals through bootstrapping.

The BLR-IPCW and MLR-IPCW approaches are dependent on correct estimation of the IPCWs. Even when censoring is non-informative (i.e. happens at random), the application of weights is essential for these methods in the presence of an absorbing state. Only in the presence of informative censoring, the pseudo-values and Aalen-Johansen estimator must be calculated within subgroups of predicted risk. The BLR-IPCW and MLR-IPCW approaches require conditional independence in the reweighted population, whereas the Aalen-Johansen and pseudo-value approaches require conditional independence given the predicted risks. In the simulation, when censoring was strongly informative, we found that calibration was poor over the range of predicted risks where density was low for both methods, but there was variability in the direction and magnitude of bias, and for which states it was present. Future research may usefully compare these two ways of dealing with informative censoring. Our work constitutes phase I and phase II of the methodological research pipeline,[80] and covers some aspects of a phase III study. Phase III and IV work comparing these methods in a wider range of scenarios would be of high value. We recommend this work focus on how the pseudo-value and IPCW approaches deal with informative censoring, with a wider range of data generating mechanisms and event rates.

This study was restricted to when the following two conditions held: 1) all individuals started in initial state, and 2) only interested in calibrating transition probabilities out of the initial state ($j = 1$) at time $t_{start} = 0$. All of the methods outlined in this paper can be implemented for $j \neq 1$ and $t_{start} > 0$ through landmarking,[46,47] where calibration would only be assessed in individuals present in state $j$ at time $t_{start}$. We recommend this should be explored in future work. While we have focused on irreversible multistate models, there is no reason these methods cannot also be applied to reversible models. Furthermore, we have focused on calibration of the transition probabilities, however there are a variety of possible approaches for assessing calibration. It may be of interest to validate the probability of a specific clinical outcome (such as developing multimorbidity in our clinical example). This risk may be the sum of multiple transition probabilities (i.e. the sum of the probabilities of being in the multimorbidity state, and of having entered death via the multimorbidity state). For this aim, graphical calibration curves[27] could be applied, where the outcome is a binary outcome: development of multimorbidity. Another approach may be to validate each competing risks sub-model for the transitions out of each state separately. This may help explore which sub-models are causing miscalibration in the estimated transition probabilities. For this aim, graphical calibration curves for competing risks models[28] would be appropriate. We focused on transition probabilities in this study as there is the least existing guidance in this area. However, we believe a study collating the methodology for each of the three approaches, and highlighting when each is appropriate, would be of use.

To our knowledge, this is the first study to detail how to assess the calibration of the transition probabilities of a multistate model, within the context of evaluating the performance of a risk prediction model. This adds to the existing literature on assessing calibration of models predicting continuous, binary, polytomous or survival outcomes.[20,21,33–36,22,26–32] All the methods considered (BLR-IPCW, MLR-IPCW and pseudo-values based on Aalen-Johansen estimator) give unbiased estimates of the calibration. We recommend producing a calibration curve using either BLR-IPCW or pseudo-values, and reporting alongside a calibration scatter plot using MLR-IPCW. Work is underway to embed the IPCW approaches into an R package.

66. The Academy of Medical Sciences. *Multimorbidity: A Priority for Global Health Research*.; 2018. https://acmedsci.ac.uk/file-download/82222577

67. Barnett K, Mercer SW, Norbury M, Watt G, Wyke S, Guthrie B. Epidemiology of multimorbidity and implications for health care, research, and medical education: A cross-sectional study. *Lancet*. 2012;380(9836):37-43. doi:10.1016/S0140-6736(12)60240-2

68. Blodgett JM, Rockwood K, Theou O. Changes in the severity and lethality of age-related health deficit accumulation in the USA between 1999 and 2018: a population-based cohort study. *Lancet Heal Longev*. 2021;2(2):e96-e104. doi:10.1016/S2666-7568(20)30059-3

69. European Observatory on Health Systems and Policies, Rijken M, Struckmann V, Dyakova M, Melchiorre MG, Al E. ICARE4EU: improving care for people with multiple chronic conditions in Europe. *Eurohealth (Lond)*. 2013;19(3):29-31.

70. Taylor AW, Price K, Gill TK, et al. Multimorbidity: not just an older person's issue. *BMC Public Health*. 2010;10(718):1-10. https://bmcpublichealth.biomedcentral.com/track/pdf/10.1186/1471-2458-10-718

71. Afshar S, Roderick PJ, Kowal P, Dimitrov BD, Hill AG. Multimorbidity and the inequalities of global ageing: A cross-sectional study of 28 countries using the World Health Surveys. *BMC Public Health*. 2015;15(1):1-10. doi:10.1186/s12889-015-2008-7

72. Arokiasamy P, Uttamacharya U, Jain K, et al. The impact of multimorbidity on adult physical and mental health in low- and middle-income countries: What does the study on global ageing and adult health (SAGE) reveal? *BMC Med*. 2015;13(1):1-16. doi:10.1186/s12916-015-0402-8

73. Garin N, Koyanagi A, Chatterji S, et al. Global Multimorbidity Patterns: A Cross-Sectional, Population-Based, Multi-Country Study. *Journals Gerontol - Ser A Biol Sci Med Sci*. 2016;71(2):205-214. doi:10.1093/gerona/glv128

74. Einarson TR, Acs A, Ludwig C, Panton UH. Economic Burden of Cardiovascular Disease in Type 2 Diabetes: A Systematic Review. *Value Heal*. 2018;21(7):881-890. doi:10.1016/j.jval.2017.12.019

75. Nichols GA, Brown JB. The impact of cardiovascular disease on medical care costs in subjects with and without type 2 diabetes. *Diabetes Care*. 2002;25(3):482-486. doi:10.2337/diacare.25.3.482

76. Wolf A, Dedman D, Campbell J, et al. Data resource profile: Clinical Practice Research Datalink (CPRD) Aurum. *Int J Epidemiol*. 2019;48(6):1740-1740G. doi:10.1093/ije/dyz034

77. Shu X, Schaubel DE. Methods for Contrasting Gap Time Hazard Functions: Application to Repeat Liver Transplantation. *Stat Biosci*. 2017;9(2):470-488. doi:10.1007/s12561-016-9168-6

78. Dabrowska D. Estimation of transition probabilities and bootstrap in a semiparametric markov renewal model. *J Nonparametr Stat*. 1995;5(3):237-259. doi:10.1080/10485259508832646

79. Spitoni C, Verduijn M, Putter H. Estimation and asymptotic theory for transition probabilities in markov renewal multi-state models. *Int J Biostat*. 2014;8(1). doi:10.1515/1557-4679.1375

80. Heinze G, Boulesteix A-L, Kammer M, Morris TP, White IR. Phases of methodological research in biostatistics - building the evidence base for new methods. 2022;(December 2022):1-8. doi:10.1002/bimj.202200222

81. RStudio: Integrated Development for R. RStudio Team. Published online 2020.

# 7. Supporting Statements

## 7.1. Acknowledgements

*The authors would like to thank the Research IT team for their assistance and the use of the Computational Shared Facility at The University of Manchester,* on which all the simulations were run.

## 7.2. Funding

This work was supported by funding from the MRC-NIHR Methodology Research Programme [grant number: MR/T025085/1]. NIHR supports IB as Senior Investigator.

## 7.3. Availability of data and materials

The simulation was implemented in R version 4.1.2,[32] and rstudio[81] using the following packages:[20,49,62,82–94]. Code is available from our GitHub public repository.[64] CPRD data must be obtained directly from the provider.

## 7.4. Author contributions

**AP and GM** conceived and designed the study in discussion with **MS**, **RR, IB** and **NP**. **AP** conducted the analysis and interpreted the results in discussion with **all authors**. **AP** wrote the initial draft of the manuscript with support from **GM**, which was then critically reviewed for important intellectual content by **all authors**. All authors have approved the final version of the paper.

## 7.5. Competing interests

None

## 7.6. Ethics approval

Access to CPRD data is supported by ISAC protocol 20_000102.

## 7.7. Supporting web materials

Supplementary material 1 – further methodology and supplementary tables

Supplementary material 2 – supplementary figures

Supplementary material 2 file could not be uploaded as the file size was too large for the manuscript submission system. We have therefore provided a link to the location of the supplementary material on GitHub. We apologise for any inconvenience.

https://github.com/manchester-predictive-healthcare-group/CHI-MRC-multi-outcome/tree/main/Project%206%20Calibration%20plots%20for%20multistate%20risk%20prediction%20models/manuscript%20files

# Supplementary material part 1 – further methodology and supplementary tables

**Associated manuscript:** Calibration plots for multistate risk prediction models: an overview and simulation comparing novel approaches

## Contents

# 8. Discussion of reliance on the Aalen-Johansen estimator on the Markov assumption in this setting

In general, the Aalen-Johansen estimator is only a consistent estimator under the Markov assumption for the multistate process $X(t)$, which means at all times $t$ "*given the present state and the event history of a patient, the next state to be visited and the time at which this will occur will only depend on the present state*".[1] If this does not hold, the estimator may give a biased estimate of transition probabilities for non-Markov processes (and in turn a biased estimator of calibration), however this is not the case for this specific setting. Datta and Satten[2] and Glidden[3] show that the Aalen-Johansen estimator is a consistent estimator of the state occupation probabilities. State occupation probabilities are defined as:

$$\boldsymbol{P}(t) = \boldsymbol{\pi}(0).\boldsymbol{p}(0,t),$$

where $\boldsymbol{\pi}(0)$ is the vector of probabilities of being in each state at $t = 0$, such that $\pi_k(0) = P(X(0) = k)$, and $\boldsymbol{p}(0,t)$ is the $K$x$K$ matrix of transition probabilities where element $\{j, k\}$ is $p_{j,k}(0,t)$.

When everyone starts in the initial state, all elements of $\boldsymbol{\pi}(0) = 0$ except the first element, $\pi_1(0) = 1$. The state occupational probabilities then reduce to the vector of transitions probabilities out of state 1:

$$\boldsymbol{P}(t) = \left(p_{1,1}(0,t), p_{1,2}(0,t), \dots, p_{1,K}(0,t)\right)$$

which is what we are trying to estimate.

Therefore, by the work of Datta and Satten[2] and Glidden,[3] the Aalen-Johansen estimator gives a consistent estimate of the transition probabilities out of state 1, and the process outlined at the start of this section will give an unbiased estimate of calibration even for non-Markov multistate processes. In summary, if all individuals start in the same initial state, and interest only lies in estimating transition probabilities out of the initial state, the Aalen-Johansen estimator is sufficient for assessing calibration. Estimators which do not depend on the Markov assumption[4–7] are not required.

# 9. Formulas to calculate the true transition probabilities under data generating mechanism 1

## 9.1. Preliminaries, notation and overview of how transition probabilities are calculated

The cause specific hazards for transitions from state $i$ to state $j$ at time $t$, $\lambda_{ij}(t)$ are $\lambda_{12}, \lambda_{13}, \lambda_{15}, \lambda_{24}, \lambda_{25}, \lambda_{34}, \lambda_{35}, \lambda_{45}$.

Let $S_i(t)$ be the survival function for state $i$, the probability of not having suffered any of the competing events out of state $i$ prior to time $t$ (assuming one is in state $i$ at time point 0). $S_i(t) = \exp(-\sum_j cumhaz_j(t))$, where $cumhaz_j(t)$ is the cumulative cause-specific hazard up to time $t$, for transition to state $j$.

Let $P_{ij}(u, t)$ be the conditional probability of being in state $j$ at time $t$, when in state $i$ at time $u$.

Let $P_{ij}^{route}(u, t)$ be the conditional probability of being in state $j$ at time $t$, when in state $i$ at time $u$, and having reached state $j$ via the specified route, $route$. Note that here $route$ could be a collection of states (for example reaching state 5 from state 1, via states 2 and 4, would be denoted $P_{25}^{24}(u, t)$

The transition probabilities we want to calculate are $P_{12}(u, t)$, $P_{13}(u, t)$, $P_{14}(u, t)$ and $P_{15}(u, t)$. Note that $P_{14}(u, t) = P_{14}^2(u, t) + P_{14}^3(u, t)$ and $P_{15}(u, t) = P_{15}^{direct}(u, t) + P_{15}^2(u, t) + P_{15}^{24}(u, t) + P_{15}^3(u, t) + P_{15}^{34}(u, t)$.

The approach is to calculate the simplest transition probabilities first (i.e. probability of transition from state 4 to state 5, $P_{45}(u, t)$, and work backwards from there. Following the theory of Putter et al.[1]

A key property used throughout, is that the probability of having not left state $i$ by time $r$, if in state $i$ at time $u$.

$$P_{ii}(u, r) = \exp(-cumhaz_i(u, r)) = \exp\left(-\left(cumhaz_i(r) - cumhaz_i(u)\right)\right) = \frac{\exp\left(-(cumhaz_i(r))\right)}{\exp\left(-(cumhaz_i(u))\right)} = \frac{S_i(r)}{S_i(u)}$$

A second key property is that the probability of being in state $j$ at time $t$, when in state $i$ at time $u$, is equal to the integral of the following quantity over all time points $r$ between $u$ and $t$: The product of (1) cause-specific hazard of transitioning out of state $i$ to some intermediate state $k$ at time $r$ ($\lambda_{ik}(r)$), (2) the probability of having remained in state $i$ until time point $r$ (see previous equation), and (3) the conditional probability of being in state $j$ after time $t$, when in state $k$ at time $r$, $P_{kj}(r, t)$.

This third quantity will have been derived by first calculating the transition probabilities at the end of the model, and using these in subsequent calculations. (i.e. first calculate the probability of transition from state 4 to state 5, $P_{45}(u,t)$, and work backwards from there).

## 9.2. Formulas for transition probabilities for data generating mechanism 1

### 9.2.1. Transition probabilities out of state 4

$$P_{44}(u,t) = \frac{S_4(t)}{S_4(u)}$$

$$P_{45}(u,t) = 1 - \frac{S_4(t)}{S_4(u)}$$

### 9.2.2. Transition probabilities out of states 2 and 3
These are symmetrical in terms of how they are derived)

**Out of state 2**

$$P_{22}(u,t) = \frac{S_2(t)}{S_2(u)}$$

$$P_{25}^{direct}(u,t) = \int_u^t \lambda_{25}(r) \frac{S_2(r)}{S_2(u)} \, dr$$

$$P_{25}^4(u,t) = \int_u^t \lambda_{24}(r) \frac{S_2(r)}{S_2(u)} P_{45}(r,t) \, dr$$

$$P_{24}(u,t) = \int_u^t \lambda_{24}(r) \frac{S_2(r)}{S_2(u)} P_{44}(r,t) \, dr$$

**Out of state 3**

$$P_{33}(u,t) = \frac{S_3(t)}{S_3(u)}$$

$$P_{35}^{direct}(u,t) = \int_u^t \lambda_{35}(r) \frac{S_3(r)}{S_3(u)} \, dr$$

$$P_{35}^4(u,t) = \int_u^t \lambda_{34}(r) \frac{S_3(r)}{S_3(u)} P_{45}(r,t) \, dr$$

$$P_{34}(u,t) = \int_u^t \lambda_{34}(r) \frac{S_3(r)}{S_3(u)} P_{44}(r,t) \, dr$$

### 9.2.3. Transition probabilities out of state 1

**Staying in state 1**

$$P_{11}(u,t) = \frac{S_1(t)}{S_1(u)}$$

**Transition to states 2 or 3**

$$P_{12}(u,t) = \int_u^t \lambda_{12}(r) \frac{S_1(r)}{S_1(u)} P_{22}(r,t)\, dr$$

$$P_{13}(u,t) = \int_u^t \lambda_{13}(r) \frac{S_1(r)}{S_1(u)} P_{33}(r,t)\, dr$$

**Transition to state 4**

$$P_{14}^2(u,t) = \int_u^t \lambda_{12}(r) \frac{S_1(r)}{S_1(u)} P_{24}(r,t)\, dr$$

$$P_{14}^3(u,t) = \int_u^t \lambda_{13}(r) \frac{S_1(r)}{S_1(u)} P_{34}(r,t)\, dr$$

$$P_{14}(u,t) = P_{14}^2(u,t) + P_{14}^3(u,t)$$

**Transition to state 5**

$$P_{15}^{direct}(u,t) = \int_u^t \lambda_{15}(r) \frac{S_1(r)}{S_1(u)}\, dr$$

$$P_{15}^2(u,t) = \int_u^t \lambda_{12}(r) \frac{S_1(r)}{S_1(u)} P_{25}^{direct}(r,t)\, dr$$

$$P_{15}^{24}(u,t) = \int_u^t \lambda_{12}(r) \frac{S_1(r)}{S_1(u)} P_{25}^4(r,t)\, dr$$

$$P_{15}^3(u,t) = \int_u^t \lambda_{13}(r) \frac{S_1(r)}{S_1(u)} P_{35}^{direct}(r,t)\, dr$$

$$P_{15}^{34}(u,t) = \int_u^t \lambda_{13}(r) \frac{S_1(r)}{S_1(u)} P_{35}^4(r,t)\, dr$$

$$P_{15}(u,t) = P_{15}^{direct}(u,t) + P_{15}^2(u,t) + P_{15}^{24}(u,t) + P_{15}^3(u,t) + P_{15}^{34}(u,t)$$

In R I will write functions to:

Calculate cause specific hazard at time t

Calculate survival functions at time t

Calculate each of the transition probabilities (will require multiple integrations I guess). Although the way it's written, each function will only have one explicitly defined integration, so it shouldn't get too messy. The others will be hidden/contained with already defined functions.

I will then generate a cohort with all individuals same predictor, and compare the event rates to the true probabilities I have derived, to see if they match.

### 9.3. Formulas for transition probabilities from DGM-2
#### 9.3.1. Transition probabilities out of states 4 and 5
They are symmetrical in terms of how they are derived

**Out of state 5**

$$P_{55}(u,t) = \frac{S_5(t)}{S_5(u)}$$

$$P_{56}(u,t) = 1 - \frac{S_5(t)}{S_5(u)}$$

**Out of state 4**

$$P_{44}(u,t) = \frac{S_4(t)}{S_4(u)}$$

$$P_{46}(u,t) = 1 - \frac{S_4(t)}{S_4(u)}$$

### 9.4. Transition probabilities out of states 2 and 3
They are symmetrical in terms of how they are derived

**Out of state 3**

$$P_{33}(u,t) = \frac{S_3(t)}{S_3(u)}$$

$$P_{36}^{direct}(u,t) = \int_u^t \lambda_{36}(r) \frac{S_3(r)}{S_3(u)} \, dr$$

$$P_{36}^5(u,t) = \int_u^t \lambda_{35}(r) \frac{S_3(r)}{S_3(u)} P_{56}(r,t) \, dr$$

$$P_{35}(u,t) = \int_u^t \lambda_{35}(r) \frac{S_3(r)}{S_3(u)} P_{55}(r,t)\, dr$$

**Out of state 2**

$$P_{22}(u,t) = \frac{S_2(t)}{S_2(u)}$$

$$P_{26}^{direct}(u,t) = \int_u^t \lambda_{26}(r) \frac{S_2(r)}{S_2(u)}\, dr$$

$$P_{26}^4(u,t) = \int_u^t \lambda_{24}(r) \frac{S_2(r)}{S_2(u)} P_{46}(r,t)\, dr$$

$$P_{24}(u,t) = \int_u^t \lambda_{24}(r) \frac{S_2(r)}{S_2(u)} P_{44}(r,t)\, dr$$

### 9.4.1. Transition probabilities out of state 1
**Staying in state 1**

$$P_{11}(u,t) = \frac{S_1(t)}{S_1(u)}$$

**Transition to states 2 or 3**

$$P_{12}(u,t) = \int_u^t \lambda_{12}(r) \frac{S_1(r)}{S_1(u)} P_{22}(r,t)\, dr$$

$$P_{13}(u,t) = \int_u^t \lambda_{13}(r) \frac{S_1(r)}{S_1(u)} P_{33}(r,t)\, dr$$

**Transition to state 4**

$$P_{14}(u,t) = P_{14}^2(u,t) = \int_u^t \lambda_{12}(r) \frac{S_1(r)}{S_1(u)} P_{24}(r,t)\, dr$$

**Transition to state 5**

$$P_{15}(u,t) = P_{15}^3(u,t) = \int_u^t \lambda_{13}(r) \frac{S_1(r)}{S_1(u)} P_{35}(r,t)\, dr$$

**Transition to state 6**

$$P_{16}^{direct}(u,t) = \int_u^t \lambda_{16}(r)\frac{S_1(r)}{S_1(u)}\,dr$$

$$P_{16}^2(u,t) = \int_u^t \lambda_{12}(r)\frac{S_1(r)}{S_1(u)} P_{26}^{direct}(r,t)\,dr$$

$$P_{16}^{24}(u,t) = \int_u^t \lambda_{12}(r)\frac{S_1(r)}{S_1(u)} P_{26}^4(r,t)\,dr$$

$$P_{16}^3(u,t) = \int_u^t \lambda_{13}(r)\frac{S_1(r)}{S_1(u)} P_{36}^{direct}(r,t)\,dr$$

$$P_{16}^{35}(u,t) = \int_u^t \lambda_{13}(r)\frac{S_1(r)}{S_1(u)} P_{36}^5(r,t)\,dr$$

$$P_{15}(u,t) = P_{16}^{direct}(u,t) + P_{16}^2(u,t) + P_{16}^{24}(u,t) + P_{16}^3(u,t) + P_{16}^{35}(u,t)$$

# 10. Code lists and operational definitions for variables in clinical example

## 10.1. Operational definitions for extracting variables

The index date is defined as the start of follow up: maximum of date turned age 65, 1st Jan 2000, and date of 1 year of up to standard registration in the database

For predictors which are 'history of', we derived a variable which indicates whether an individual has a record of the comorbidity prior to their index date in their primary care record.

For predictors reliant on test data (BMI, SBP, cholesterol/HDL ratio, smoking status), we looked in the five years prior to the index date for an occurrence of the variable. Appropriate conversions were applied based on the unit of measurement recorded in the database. Extreme values were then removed. See Github page for full algorithms for extraction of each variable.[8]

For outcomes, we derived the time until first occurrence or censoring, and a censoring indicator, in both the primary care, secondary care and ONS datasets. We then took the event as the time until the first of any of these to occur. We also derived presence of each condition at baseline, in order to apply our exclusion criteria. Cardiovascular disease was a composite variable consisting of coronary heart disease, heart failure, myocardial infarction, stroke and transient ischaemic attack.

CKD (stage 3, 4 or 5) events were identified through medical codes and test data, namely eGFR scores and creatinine measurements. The process for doing so is below.

**Identifying CKD from eGFR/GFR, and estimating eGFR/GFR from creatinine**

KDIGO guidelines give definition of CKD.[9] This is from 2012, but are currently still the guidelines they recommend. Abnormalities of kidney function must be present for > 3 months. This is CKD-EPI equation[10] they recommend for converting creatinine mesaurements to GFR/eGFR scores (also requires sex, ethnicity and age). Note that they state something along the lines of "this, or any equation shown to be better than this" can be used. Therefore there may be a more recent equation for this conversion, as the one they recommend was developed in 2009. This comparison shows the CKD-EPI equation is better than the MDRD, which is another alternative.[11]

The attached code looks for two entries below a certain value (60) that are more than 90 days (3 months) apart. Dataset should be in format of:

- 1 row per eGFR score
- Variables: person_id(identifier for individuals), EntryDate (date of code), CodeValue (value of eGFR), num (increasing integer indicating the observation number for each individual)

algorithm_convert_gfr.R

**Overall strategy:**

1) Identify CKD stages 3/4/5 using medical codes
2) Extract eGFR/GFR + creatinine scores from test data
3) Convert creatinine measurements to eGFR scores
4) Use algorithm to identify if individuals meets criteria for CKD stages 3/4/5 using algorithm provided

5) CKD = 1 if either medical code, or CKD identified from test data.

## 10.2. Code lists used for primary care extraction

All codelists are available on GitHub.[8] Names of the final codelists used for data extraction, and details of where these were obtained is given in this section.

For this study, we opt to focus only on medical codes, and ignore prescription data. Table SM2 contains all the names of the code lists used for extraction of the primary care data and the source from which they were obtained. Code lists from source "AH" are available at the referenced Github page.[12,13] Code lists from source "Cambridge mapped" were taken from the Cambridge primary care unit website,[14] and then mapped by author TVS from CPRD GOLD to CPRD Aurum. Ethnicity code list was available at on the LSHTM Data Compass.[15,16] The groupings for the medical codes were obtained from a Read code code list on another GitHub respository[17], developed for another study.[18] Some additional ethnicity codes were identified from this second code list, found in the CPRD Aurum code browser, and added to the code list. The smoking status code list was available on the LSHTM Data Compass.[19,20] Self-generated code lists we developed by searching the CPRD Aurum code browser for *height*, *weight*, *body mass*, *systolic* and *cholesterol* respectively, and selecting the appropriate codes. Finally, we opted to use code lists for Psoriasis (as opposed to Psoriasis and eczema) and Rheumatoid arthritis (as opposed to rheumatoid arthritis and connective tissue disorders), due to the availability of code lists from source AH.

*Table SM2: Code lists file names and source*

|   | Variable | Code list name | Source |
|---|---|---|---|
| | **Medical history** | | |
| 1 | Alcohol misuse | Alcohol_misuse | AH |
| 5 | Depression | Depression | AH |
| 9 | Chronic kidney disease | Chronic_kidney_disease | AH |
| 13 | Coronary heart disease | Coronary_heart_disease | AH |
| 15 | Diabetes | Diabetes_mellitus_other_or_not_specified | AH |
| 15 | Diabetes | Diabetes_mellitus_type1 | AH |
| 15 | Diabetes | Diabetes_mellitus_type2 | AH |
| 19 | Heart failure | Heart_failure | AH |
| 20 | Hypertension | Hypertension | AH |
| 33 | Stroke transient ischaemic attack | Stroke_not_otherwise_specified | AH |
| 33 | Stroke transient ischaemic attack | Transient_ischaemic_attack | AH |
| 35 | Myocardial Infarction | Myocardial_infarction | AH |
| | **Demographic, lifestyle and test data** | | |
| 37 | Body mass index | height | Self-generated |
| 37 | Body mass index | weight | Self-generated |
| 37 | Body mass index | bmi | Self-generated |
| 38 | Systolic blood pressure | sbp | Self-generated |
| 39 | Smoking status | cr_smokingcodes_aurum | Internet search |

| | Cholesterol | chol_total | Self-generated |
| --- | --- | --- | --- |
| 40 | | | |
| 40 | Cholesterol | chol_hdl | Self-generated |
| 40 | Cholesterol | chol_ratio | Self-generated |

Note that some of the code lists were edited after being moved onto incline as some formatting changes had to be made. I am keeping a table of the pre-formatted variable names for personal reference (Table 2) as both have been saved on GitHub.[8]

*Table SM3: Pre-formatted code list names and post-formatted code list names*

| Variable | Pre-format name | Post-format name |
| --- | --- | --- |
| Inflammatory bowel disease | IBD160_mapped | IBD160_mcid |
| Peptic Ulcer Disease | PEP135_mapped | PEP135_mcid |
| Prostate disorders | PRO170_mapped | PRO170_mcid |
| Ethnicity | Ethnicity_aurum_wgroups | Ethnicity_aurum_wgroups_mcid |
| Body mass index | height | Height_mcid |
| Body mass index | weight | Weight_mcid |
| Body mass index | bmi | bmi_mcid |
| Systolic blood pressure | sbp | sbp_mcid |
| Smoking status | cr_smokingcodes_aurum | cr_smokingcodes_aurum_mcid |
| Cholesterol | chol_total | chol_total_mcid |
| Cholesterol | chol_hdl | chol_hdl_mcid |
| Cholesterol | chol_ratio | chol_ratio_mcid |

### 10.3. Code lists used for HES/ONS extraction

Table SM4 contains the variables and ICD 10 codes used to extract it. For chronic kidney disease, the full 5 digit ICD 10 code is required to separate CKD stages 1 and 2 from stages 3, 4 and 5. All others only require the initial 3 digits for extraction. Following the process in QRISK3, CHD and MI were grouped, and Stroke and TIA were grouped into one outcome.

Table SM4: ICD 10 codes used for extraction

| Variable | ICD 10 codes | # characters |
| --- | --- | --- |
| AF | 'I48' | 3 |
| CHD/MI | 'I20','I21','I22','I23','I24','I25' | 3 |

| Stroke/TIA | 'G45','I63','I64' | 3 |
| Heart failure | 'I50' | 3 |
| Type 1 diabetes | 'E10' | 3 |
| Type 2 diabetes | 'E11' | 3 |
| Chronic kidney disease stage 3/4/5 | 'N18.3','N18.30','N18.31','N18.32','N18.4','N18.5','N18.6','N18.9' | 5 |

72

73

## 11. Supplementary Tables

*Table S1: Number of transitions in and out of each state in the clinical example*

| | | Development dataset (N = 5,000) | | | | | | | |
|---|---|---|---|---|---|---|---|---|---|
| | | Entering | | | | | | | |
| | | State 1 | State 2 | State 3 | State 4 | State 5 | State 6 | no event | total entering |
| Leaving | State 1 | 0 | 1002 | 629 | 566 | 14 | 530 | 2259 | 5000 |
| | State 2 | 0 | 0 | 0 | 0 | 427 | 250 | 325 | 1002 |
| | State 3 | 0 | 0 | 0 | 0 | 259 | 69 | 301 | 629 |
| | State 4 | 0 | 0 | 0 | 0 | 148 | 124 | 294 | 566 |
| | State 5 | 0 | 0 | 0 | 0 | 0 | 379 | 469 | 848 |
| | State 6 | 0 | 0 | 0 | 0 | 0 | 0 | 1352 | 1352 |
| | | Validation dataset (N = 100,000) | | | | | | | |
| | | Entering | | | | | | | |
| | | State 1 | State 2 | State 3 | State 4 | State 5 | State 6 | no event | total entering |
| Leaving | State 1 | 0 | 19299 | 12891 | 12091 | 325 | 10854 | 44540 | 100000 |
| | State 2 | 0 | 0 | 0 | 0 | 7650 | 5165 | 6484 | 19299 |
| | State 3 | 0 | 0 | 0 | 0 | 5517 | 1611 | 5763 | 12891 |
| | State 4 | 0 | 0 | 0 | 0 | 3189 | 2524 | 6378 | 12091 |
| | State 5 | 0 | 0 | 0 | 0 | 0 | 7424 | 9257 | 16681 |
| | State 6 | 0 | 0 | 0 | 0 | 0 | 0 | 27578 | 27578 |

# Supplementary material part 2 – supplementary figures

**Associated manuscript: Calibration plots for multistate risk prediction models: an overview and simulation comparing novel approaches**

Supplementary material 2 file could not be uploaded as the file size was too large for the manuscript submission system. We have therefore provided a link to the location of the supplementary material on GitHub. We apologise for any inconvenience.

https://github.com/manchester-predictive-healthcare-group/CHI-MRC-multi-outcome/tree/main/Project%206%20Calibration%20plots%20for%20multistate%20risk%20prediction%20models/manuscript%20files